 \documentclass[aps,prl,twocolumn,showpacs,superscriptaddress,groupedaddress]{revtex4}  % for review and submission
\usepackage{graphicx}  % needed for figures
\usepackage{dcolumn}   % needed for some tables
\usepackage{bm}        % for math
\usepackage{amssymb}   % for math
\usepackage{amsmath}
\usepackage{color}
\usepackage{multirow}
 \usepackage{rotating}
\usepackage{xspace}
\usepackage{comment}
\usepackage{slashed}

\newcommand{\fix}[1]{} % comment this in if you want to the see the PRL without comments
\newcommand{\MET}{\mbox{$\raisebox{.3ex}{$\not\!$}E_T$}}

\newcommand{\ttbar}     {\mbox{$t\bar{t}$}\xspace}

\newcommand{\ppbar}     {\mbox{$p\bar{p}$}\xspace}
\newcommand{\ljets}     {\mbox{$\ell$+jets}\xspace}
\newcommand{\comphep}   {\sc comphep}
\newcommand{\singletop} {\sc singletop}
\newcommand{\pythia}    {\mbox{\textsc{pythia}}}

\newcommand{\geant}     {{\sc{geant}}}
\newcommand{\alpgen}    {\mbox{\textsc{alpgen}}}
\newcommand{\mcfm}      {\sc mcfm}

\newcommand{\ifb}       {fb$^{-1}$}

\newcommand{\Wt}{\ensuremath{\mathit{\!Wt}}\xspace} 

\newcommand{\xsectev}{1.29}
\newcommand{\xsecteverrorup}{+0.26}
\newcommand{\xsecteverrordown}{-0.24}

\begin{document}

\hspace{5.2in} \mbox{FERMILAB-PUB-15-088-E}

\title{Tevatron Combination of Single-Top-Quark Cross Sections and
  Determination of the Magnitude of the Cabibbo-Kobayashi-Maskawa
  Matrix Element $\bf V_{tb}$}

%%%%%%%%%%%%%%%%%%%%%%%%%%%
%
\affiliation{Institute of Physics, Academia Sinica, Taipei, Taiwan 11529, Republic of China}
\affiliation{Argonne National Laboratory, Argonne, Illinois 60439, USA}
\affiliation{University of Athens, 157 71 Athens, Greece}
\affiliation{Institut de Fisica d'Altes Energies, ICREA, Universitat Autonoma de Barcelona, E-08193, Bellaterra (Barcelona), Spain}
\affiliation{Baylor University, Waco, Texas 76798, USA}
\affiliation{Istituto Nazionale di Fisica Nucleare Bologna, \ensuremath{^{yy}}University of Bologna, I-40127 Bologna, Italy}
\affiliation{University of California, Davis, Davis, California 95616, USA}
\affiliation{University of California, Los Angeles, Los Angeles, California 90024, USA}
\affiliation{Instituto de Fisica de Cantabria, CSIC-University of Cantabria, 39005 Santander, Spain}
\affiliation{Carnegie Mellon University, Pittsburgh, Pennsylvania 15213, USA}
\affiliation{Enrico Fermi Institute, University of Chicago, Chicago, Illinois 60637, USA}
\affiliation{Comenius University, 842 48 Bratislava, Slovakia; Institute of Experimental Physics, 040 01 Kosice, Slovakia}
\affiliation{Joint Institute for Nuclear Research, RU-141980 Dubna, Russia}
\affiliation{Duke University, Durham, North Carolina 27708, USA}
\affiliation{Fermi National Accelerator Laboratory, Batavia, Illinois 60510, USA}
\affiliation{University of Florida, Gainesville, Florida 32611, USA}
\affiliation{Laboratori Nazionali di Frascati, Istituto Nazionale di Fisica Nucleare, I-00044 Frascati, Italy}
\affiliation{University of Geneva, CH-1211 Geneva 4, Switzerland}
\affiliation{Glasgow University, Glasgow G12 8QQ, United Kingdom}
\affiliation{Harvard University, Cambridge, Massachusetts 02138, USA}
\affiliation{Division of High Energy Physics, Department of Physics, University of Helsinki, FIN-00014, Helsinki, Finland; Helsinki Institute of Physics, FIN-00014, Helsinki, Finland}
\affiliation{University of Illinois, Urbana, Illinois 61801, USA}
\affiliation{The Johns Hopkins University, Baltimore, Maryland 21218, USA}
\affiliation{Institut f\"{u}r Experimentelle Kernphysik, Karlsruhe Institute of Technology, D-76131 Karlsruhe, Germany}
\affiliation{Center for High Energy Physics: Kyungpook National University, Daegu 702-701, Korea; Seoul National University, Seoul 151-742, Korea; Sungkyunkwan University, Suwon 440-746, Korea; Korea Institute of Science and Technology Information, Daejeon 305-806, Korea; Chonnam National University, Gwangju 500-757, Korea; Chonbuk National University, Jeonju 561-756, Korea; Ewha Womans University, Seoul, 120-750, Korea}
\affiliation{Ernest Orlando Lawrence Berkeley National Laboratory, Berkeley, California 94720, USA}
\affiliation{University of Liverpool, Liverpool L69 7ZE, United Kingdom}
\affiliation{University College London, London WC1E 6BT, United Kingdom}
\affiliation{Centro de Investigaciones Energeticas Medioambientales y Tecnologicas, E-28040 Madrid, Spain}
\affiliation{Massachusetts Institute of Technology, Cambridge, Massachusetts 02139, USA}
\affiliation{University of Michigan, Ann Arbor, Michigan 48109, USA}
\affiliation{Michigan State University, East Lansing, Michigan 48824, USA}
\affiliation{Institution for Theoretical and Experimental Physics, ITEP, Moscow 117259, Russia}
\affiliation{University of New Mexico, Albuquerque, New Mexico 87131, USA}
\affiliation{The Ohio State University, Columbus, Ohio 43210, USA}
\affiliation{Okayama University, Okayama 700-8530, Japan}
\affiliation{Osaka City University, Osaka 558-8585, Japan}
\affiliation{University of Oxford, Oxford OX1 3RH, United Kingdom}
\affiliation{Istituto Nazionale di Fisica Nucleare, Sezione di Padova, \ensuremath{^{zz}}University of Padova, I-35131 Padova, Italy}
\affiliation{University of Pennsylvania, Philadelphia, Pennsylvania 19104, USA}
\affiliation{Istituto Nazionale di Fisica Nucleare Pisa, \ensuremath{^{aaa}}University of Pisa, \ensuremath{^{bbb}}University of Siena, \ensuremath{^{ccc}}Scuola Normale Superiore, I-56127 Pisa, Italy, \ensuremath{^{ddd}}INFN Pavia, I-27100 Pavia, Italy, \ensuremath{^{eee}}University of Pavia, I-27100 Pavia, Italy}
\affiliation{University of Pittsburgh, Pittsburgh, Pennsylvania 15260, USA}
\affiliation{Purdue University, West Lafayette, Indiana 47907, USA}
\affiliation{University of Rochester, Rochester, New York 14627, USA}
\affiliation{The Rockefeller University, New York, New York 10065, USA}
\affiliation{Istituto Nazionale di Fisica Nucleare, Sezione di Roma 1, \ensuremath{^{fff}}Sapienza Universit\`{a} di Roma, I-00185 Roma, Italy}
\affiliation{Mitchell Institute for Fundamental Physics and Astronomy, Texas A\&M University, College Station, Texas 77843, USA}
\affiliation{Istituto Nazionale di Fisica Nucleare Trieste, \ensuremath{^{ggg}}Gruppo Collegato di Udine, \ensuremath{^{hhh}}University of Udine, I-33100 Udine, Italy, \ensuremath{^{iii}}University of Trieste, I-34127 Trieste, Italy}
\affiliation{University of Tsukuba, Tsukuba, Ibaraki 305, Japan}
\affiliation{Tufts University, Medford, Massachusetts 02155, USA}
\affiliation{University of Virginia, Charlottesville, Virginia 22906, USA}
\affiliation{Waseda University, Tokyo 169, Japan}
\affiliation{Wayne State University, Detroit, Michigan 48201, USA}
\affiliation{University of Wisconsin, Madison, Wisconsin 53706, USA}
\affiliation{Yale University, New Haven, Connecticut 06520, USA}
\affiliation{LAFEX, Centro Brasileiro de Pesquisas F\'{i}sicas, Rio de Janeiro, Brazil}
\affiliation{Universidade do Estado do Rio de Janeiro, Rio de Janeiro, Brazil}
\affiliation{Universidade Federal do ABC, Santo Andr\'{e}, Brazil}
\affiliation{University of Science and Technology of China, Hefei, People's Republic of China}
\affiliation{Universidad de los Andes, Bogot\'{a}, Colombia}
\affiliation{Charles University, Faculty of Mathematics and Physics, Center for Particle Physics, Prague, Czech Republic}
\affiliation{Czech Technical University in Prague, Prague, Czech Republic}
\affiliation{Institute of Physics, Academy of Sciences of the Czech Republic, Prague, Czech Republic}
\affiliation{Universidad San Francisco de Quito, Quito, Ecuador}
\affiliation{LPC, Universit\'{e} Blaise Pascal, CNRS/IN2P3, Clermont, France}
\affiliation{LPSC, Universit\'{e} Joseph Fourier Grenoble 1, CNRS/IN2P3, Institut National Polytechnique de Grenoble, Grenoble, France}
\affiliation{CPPM, Aix-Marseille Universit\'{e}, CNRS/IN2P3, Marseille, France}
\affiliation{LAL, Universit\'{e} Paris-Sud, CNRS/IN2P3, Orsay, France}
\affiliation{LPNHE, Universit\'{e}s Paris VI and VII, CNRS/IN2P3, Paris, France}
\affiliation{CEA, Irfu, SPP, Saclay, France}
\affiliation{IPHC, Universit\'{e} de Strasbourg, CNRS/IN2P3, Strasbourg, France}
\affiliation{IPNL, Universit\'{e} Lyon 1, CNRS/IN2P3, Villeurbanne, France and Universit\'{e} de Lyon, Lyon, France}
\affiliation{III. Physikalisches Institut A, RWTH Aachen University, Aachen, Germany}
\affiliation{Physikalisches Institut, Universit\"{a}t Freiburg, Freiburg, Germany}
\affiliation{II. Physikalisches Institut, Georg-August-Universit\"{a}t G\"{o}ttingen, G\"{o}ttingen, Germany}
\affiliation{Institut f\"{u}r Physik, Universit\"{a}t Mainz, Mainz, Germany}
\affiliation{Ludwig-Maximilians-Universit\"{a}t M\"{u}nchen, M\"{u}nchen, Germany}
\affiliation{Panjab University, Chandigarh, India}
\affiliation{Delhi University, Delhi, India}
\affiliation{Tata Institute of Fundamental Research, Mumbai, India}
\affiliation{University College Dublin, Dublin, Ireland}
\affiliation{Korea Detector Laboratory, Korea University, Seoul, Korea}
\affiliation{CINVESTAV, Mexico City, Mexico}
\affiliation{Nikhef, Science Park, Amsterdam, the Netherlands}
\affiliation{Radboud University Nijmegen, Nijmegen, the Netherlands}
\affiliation{Joint Institute for Nuclear Research, RU-141980 Dubna, Russia}
\affiliation{Institution for Theoretical and Experimental Physics, ITEP, Moscow 117259, Russia}
\affiliation{Moscow State University, Moscow, Russia}
\affiliation{Institute for High Energy Physics, Protvino, Russia}
\affiliation{Petersburg Nuclear Physics Institute, St. Petersburg, Russia}
\affiliation{Instituci\'{o} Catalana de Recerca i Estudis Avan\c{c}ats (ICREA) and Institut de F\'{i}sica d'Altes Energies (IFAE), Barcelona, Spain}
\affiliation{Uppsala University, Uppsala, Sweden}
\affiliation{Taras Shevchenko National University of Kyiv, Kiev, Ukraine}
\affiliation{Lancaster University, Lancaster LA1 4YB, United Kingdom}
\affiliation{Imperial College London, London SW7 2AZ, United Kingdom}
\affiliation{The University of Manchester, Manchester M13 9PL, United Kingdom}
\affiliation{University of Arizona, Tucson, Arizona 85721, USA}
\affiliation{University of California Riverside, Riverside, California 92521, USA}
\affiliation{Florida State University, Tallahassee, Florida 32306, USA}
\affiliation{Fermi National Accelerator Laboratory, Batavia, Illinois 60510, USA}
\affiliation{University of Illinois at Chicago, Chicago, Illinois 60607, USA}
\affiliation{Northern Illinois University, DeKalb, Illinois 60115, USA}
\affiliation{Northwestern University, Evanston, Illinois 60208, USA}
\affiliation{Indiana University, Bloomington, Indiana 47405, USA}
\affiliation{Purdue University Calumet, Hammond, Indiana 46323, USA}
\affiliation{University of Notre Dame, Notre Dame, Indiana 46556, USA}
\affiliation{Iowa State University, Ames, Iowa 50011, USA}
\affiliation{University of Kansas, Lawrence, Kansas 66045, USA}
\affiliation{Louisiana Tech University, Ruston, Louisiana 71272, USA}
\affiliation{Northeastern University, Boston, Massachusetts 02115, USA}
\affiliation{University of Michigan, Ann Arbor, Michigan 48109, USA}
\affiliation{Michigan State University, East Lansing, Michigan 48824, USA}
\affiliation{University of Mississippi, University, Mississippi 38677, USA}
\affiliation{University of Nebraska, Lincoln, Nebraska 68588, USA}
\affiliation{Rutgers University, Piscataway, New Jersey 08855, USA}
\affiliation{Princeton University, Princeton, New Jersey 08544, USA}
\affiliation{State University of New York, Buffalo, New York 14260, USA}
\affiliation{University of Rochester, Rochester, New York 14627, USA}
\affiliation{State University of New York, Stony Brook, New York 11794, USA}
\affiliation{Brookhaven National Laboratory, Upton, New York 11973, USA}
\affiliation{Langston University, Langston, Oklahoma 73050, USA}
\affiliation{University of Oklahoma, Norman, Oklahoma 73019, USA}
\affiliation{Oklahoma State University, Stillwater, Oklahoma 74078, USA}
\affiliation{Brown University, Providence, Rhode Island 02912, USA}
\affiliation{University of Texas, Arlington, Texas 76019, USA}
\affiliation{Southern Methodist University, Dallas, Texas 75275, USA}
\affiliation{Rice University, Houston, Texas 77005, USA}
\affiliation{University of Virginia, Charlottesville, Virginia 22904, USA}
\affiliation{University of Washington, Seattle, Washington 98195, USA}

\author{T.~Aaltonen \ensuremath{^{\dagger}}}
\affiliation{Division of High Energy Physics, Department of Physics, University of Helsinki, FIN-00014, Helsinki, Finland; Helsinki Institute of Physics, FIN-00014, Helsinki, Finland}
\author{V.M.~Abazov \ensuremath{^{\ddagger}}}
\affiliation{Joint Institute for Nuclear Research, RU-141980 Dubna, Russia}
\author{B.~Abbott \ensuremath{^{\ddagger}}}
\affiliation{University of Oklahoma, Norman, Oklahoma 73019, USA}
\author{B.S.~Acharya \ensuremath{^{\ddagger}}}
\affiliation{Tata Institute of Fundamental Research, Mumbai, India}
\author{M.~Adams \ensuremath{^{\ddagger}}}
\affiliation{University of Illinois at Chicago, Chicago, Illinois 60607, USA}
\author{T.~Adams \ensuremath{^{\ddagger}}}
\affiliation{Florida State University, Tallahassee, Florida 32306, USA}
\author{J.P.~Agnew \ensuremath{^{\ddagger}}}
\affiliation{The University of Manchester, Manchester M13 9PL, United Kingdom}
\author{G.D.~Alexeev \ensuremath{^{\ddagger}}}
\affiliation{Joint Institute for Nuclear Research, RU-141980 Dubna, Russia}
\author{G.~Alkhazov \ensuremath{^{\ddagger}}}
\affiliation{Petersburg Nuclear Physics Institute, St. Petersburg, Russia}
\author{A.~Alton \ensuremath{^{\ddagger}}\ensuremath{^{jj}}}
\affiliation{University of Michigan, Ann Arbor, Michigan 48109, USA}
% ...

\collaboration{CDF Collaboration}
\altaffiliation[With visitors from]{
\ensuremath{^{a}}University of British Columbia, Vancouver, BC V6T 1Z1, Canada,
\ensuremath{^{b}}Istituto Nazionale di Fisica Nucleare, Sezione di Cagliari, 09042 Monserrato (Cagliari), Italy,
\ensuremath{^{c}}University of California Irvine, Irvine, CA 92697, USA,
\ensuremath{^{d}}Institute of Physics, Academy of Sciences of the Czech Republic, 182 21, Czech Republic,
\ensuremath{^{e}}CERN, CH-1211 Geneva, Switzerland,
\ensuremath{^{f}}Cornell University, Ithaca, NY 14853, USA,
\ensuremath{^{g}}University of Cyprus, Nicosia CY-1678, Cyprus,
\ensuremath{^{h}}Office of Science, U.S. Department of Energy, Washington, DC 20585, USA,
\ensuremath{^{i}}University College Dublin, Dublin 4, Ireland,
\ensuremath{^{j}}ETH, 8092 Z\"{u}rich, Switzerland,
\ensuremath{^{k}}University of Fukui, Fukui City, Fukui Prefecture, Japan 910-0017,
\ensuremath{^{l}}Universidad Iberoamericana, Lomas de Santa Fe, M\'{e}xico, C.P. 01219, Distrito Federal,
\ensuremath{^{m}}University of Iowa, Iowa City, IA 52242, USA,
\ensuremath{^{n}}Kinki University, Higashi-Osaka City, Japan 577-8502,
\ensuremath{^{o}}Kansas State University, Manhattan, KS 66506, USA,
\ensuremath{^{p}}Brookhaven National Laboratory, Upton, NY 11973, USA,
\ensuremath{^{q}}Istituto Nazionale di Fisica Nucleare, Sezione di Lecce, Via Arnesano, I-73100 Lecce, Italy,
\ensuremath{^{r}}Queen Mary, University of London, London, E1 4NS, United Kingdom,
\ensuremath{^{s}}University of Melbourne, Victoria 3010, Australia,
\ensuremath{^{t}}Muons, Inc., Batavia, IL 60510, USA,
\ensuremath{^{u}}Nagasaki Institute of Applied Science, Nagasaki 851-0193, Japan,
\ensuremath{^{v}}National Research Nuclear University, Moscow 115409, Russia,
\ensuremath{^{w}}Northwestern University, Evanston, IL 60208, USA,
\ensuremath{^{x}}University of Notre Dame, Notre Dame, IN 46556, USA,
\ensuremath{^{y}}Universidad de Oviedo, E-33007 Oviedo, Spain,
\ensuremath{^{z}}CNRS-IN2P3, Paris, F-75205 France,
\ensuremath{^{aa}}Universidad Tecnica Federico Santa Maria, 110v Valparaiso, Chile,
\ensuremath{^{bb}}The University of Jordan, Amman 11942, Jordan,
\ensuremath{^{cc}}Universite catholique de Louvain, 1348 Louvain-La-Neuve, Belgium,
\ensuremath{^{dd}}University of Z\"{u}rich, 8006 Z\"{u}rich, Switzerland,
\ensuremath{^{ee}}Massachusetts General Hospital, Boston, MA 02114 USA,
\ensuremath{^{ff}}Harvard Medical School, Boston, MA 02114 USA,
\ensuremath{^{gg}}Hampton University, Hampton, VA 23668, USA,
\ensuremath{^{hh}}Los Alamos National Laboratory, Los Alamos, NM 87544, USA,
\ensuremath{^{ii}}Universit\`{a} degli Studi di Napoli Federico I, I-80138 Napoli, Italy
}
\noaffiliation
\collaboration{D0 Collaboration}
\altaffiliation[With visitors from]{
\ensuremath{^{jj}}Augustana College, Sioux Falls, SD, USA,
\ensuremath{^{kk}}The University of Liverpool, Liverpool, UK,
\ensuremath{^{ll}}DESY, Hamburg, Germany,
\ensuremath{^{mm}}CONACyT, Mexico City, Mexico,
\ensuremath{^{nn}}Universidad Michoacana de San Nicolas de Hidalgo, Morelia, Mexico,
\ensuremath{^{oo}}SLAC, Menlo Park, CA, USA,
\ensuremath{^{pp}}University College London, London, UK,
\ensuremath{^{qq}}Centro de Investigacion en Computacion - IPN, Mexico City, Mexico,
\ensuremath{^{rr}}Universidade Estadual Paulista, S\~{a}o Paulo, Brazil,
\ensuremath{^{ss}}Karlsruher Institut f\"{u}r Technologie (KIT) - Steinbuch Centre for Computing (SCC),
\ensuremath{^{tt}}Office of Science, U.S. Department of Energy, Washington, D.C. 20585, USA,
\ensuremath{^{uu}}American Association for the Advancement of Science, Washington, D.C. 20005, USA,
\ensuremath{^{vv}}National Academy of Science of Ukraine (NASU) - Kiev Institute for Nuclear Research (KINR),
\ensuremath{^{ww}}University of Maryland, College Park, MD 20742,
\ensuremath{^{xx}}European Organization for Nuclear Research (CERN),
}
\noaffiliation  % CDF+D0 author list
%
%%%%%%%%%%%%%%%%%%%%%%%%%%%

\date{January 19, 2016}

\begin{abstract}
\noindent
We present the final combination of CDF and D0 measurements of cross
sections for single-top-quark production in proton-antiproton
collisions at a center-of-mass energy of 1.96~TeV. The data correspond
to total integrated luminosities of  up to 9.7~\ifb\ per
experiment. The $t$-channel cross section is measured to be $\sigma_t
= 2.25^{+0.29}_{-0.31}$~pb. We also present the combinations of the
two-dimensional measurements of the $s$- vs.\ $t$-channel cross
section. In addition, we give the combination of the $s+t$ channel
cross section measurement resulting 
in $\sigma_{s+t} = 3.30^{+0.52}_{-0.40}$~pb, without assuming the
standard-model value for the ratio $\sigma_s/\sigma_t$. The resulting
value of the magnitude of the top-to-bottom quark coupling is
$|V_{tb}|$ = $1.02^{+0.06}_{-0.05}$, corresponding to $|V_{tb}| >
0.92$ at the 95\% C.L.
\end{abstract}

\maketitle

%\modulolinenumbers[1]
%\linenumbers

%---------------------------------------------------------------------

%%%%%%%%%%%%%%%%%%%%%%%%%%%
%% 
%% Introduction
%% 
%%%%%%%%%%%%%%%%%%%%%%%%%%%

The top quark is the heaviest elementary particle of the standard model
(SM). Detailed studies of top-quark production and decay provide
stringent tests of strong and electroweak interactions, as well as
sensitivity to extensions of the SM~\cite{top}. At the Fermilab
Tevatron collider, protons ($p$) and antiprotons ($\bar{p}$) collided
at a center-of-mass energy of $\sqrt{s}=1.96$\;TeV. Top quarks were
produced predominantly in pairs (\ttbar) via the strong
interaction~\cite{ttbar_combi}.  They were also produced singly via
the electroweak interaction. The cross section for single-top-quark
production depends on the square of the magnitude of the quark-mixing 
Cabibbo-Kobayashi-Maskawa (CKM) matrix~\cite{ckm} element $V_{tb}$,
and consequently is sensitive to contributions from a fourth 
family of quarks~\cite{FourthGen1,FourthGen2}, as well as other 
new phenomena~\cite{Tait:2000sh}, which would lead to a measured 
strength of the $Wtb$ coupling $|V_{tb}|$ different from the SM
prediction. Non-SM phenomena could also change the relative fraction
of events produced in the various channels that contribute to the
total single-top-quark production cross section.

In $p{\bar p}$ scattering, single-top-quark production proceeds in the
$t$-channel via the exchange of a spacelike virtual $W$ boson between
a light quark and a bottom
quark~\cite{singletop-willenbrock,singletop-yuan,tchannel-kidonakis}.
Single top quarks are also produced in the $s$ channel via the decay
of a timelike virtual $W$ boson 
produced by quark-antiquark annihilation, which produces a top quark
and a bottom quark~\cite{singletop-cortese}, or in association with a
$W$ boson (\Wt)~\cite{Wt-tait}.  The predicted SM cross section for
the $t$-channel process $\sigma_t$ is $2.10 \pm
0.13$~pb~\cite{tchannel-kidonakis}, while the $s$-channel cross
section $\sigma_s$ is $1.05 \pm 0.06$~pb~\cite{schannel-kidonakis},
both calculated at next-to-leading-order (NLO) in quantum
chromodynamics (QCD), including next-to-next-to-leading logarithmic (NNLL)
corrections. A top-quark mass of 172.5~GeV was chosen, which is
consistent with the 
current world average value~\cite{topmasswa}. The cross section for
\Wt\ production $\sigma_{Wt}$  is negligibly small at the Tevatron and
 therefore is not considered in the combination described in this Letter. 
Since the magnitude of the $Wtb$ coupling is much larger than that of
$Wtd$ or of $Wts$~\cite{pdg}, each top quark decays almost exclusively
to a $W$~boson and a $b$ quark.

Observation of single-top-quark production was reported by the
CDF~\cite{stop-obs-2009-cdf,Aaltonen:2010fs,cdf-prd-2010} and
D0~\cite{stop-obs-2009-d0,stop-2011-d0} Collaborations not
differentiating between
the $s$ and the $t$ channels (hereinafter $s+t$ channel). The CDF
Collaboration subsequently measured a single-top-quark production
cross section for the sum of the $s$, $t$, and \Wt channels of
$\sigma_{s+t+\Wt} =3.04^{+0.57}_{-0.53}$~pb using data corresponding
to 7.5~\ifb\ of integrated luminosity~\cite{cdf_channels_7.5} and for
the sum of the $s$ and $t$ channels of $\sigma_{s+t} =
3.02^{+0.49}_{-0.48}$\;pb using up to 9.5~\ifb\ of integrated
luminosity~\cite{cdf_channels}. The D0 Collaboration obtained
$\sigma_{s+t} = 4.11^{+0.60}_{-0.55}$\;pb using data corresponding to
9.7~\ifb\ of integrated luminosity~\cite{d0_schannel}. 

The cross sections for individual production modes were also measured separately. 
The D0 Collaboration observed the $t$-channel
process~\cite{t-channel-new} and measured its cross section to be $\sigma_t =
3.07^{+0.54}_{-0.49}$\;pb using data corresponding to 9.7~\ifb\ of
integrated luminosity~\cite{d0_schannel}. 
The CDF Collaboration measured $\sigma_{t+\Wt} =
1.66^{+0.53}_{-0.47}$~pb using data corresponding to 7.5~\ifb\ of
integrated luminosity~\cite{cdf_channels_7.5} and $\sigma_t =
1.65^{+0.38}_{-0.36}$\;pb using up to 9.5~\ifb~\cite{cdf_channels} of
integrated luminosity. The difference between the results for
$\sigma_t$ is about 2 standard deviations (s.\ d.). 
Furthermore, both the CDF and D0 Collaborations reported evidence for
$s$-channel
production~\cite{cdf_schannel,cdf_schannel_MET,d0_schannel} and
combined their results 
to observe the $s$-channel process with $\sigma_s =
\xsectev^{\xsecteverrorup}_{\xsecteverrordown}$~pb~\cite{tev_schannel}. 

At the CERN LHC proton-proton ($pp$) collider, $t$-channel production
was observed by the ATLAS and CMS
Collaborations~\cite{atlas-tchannel-1,atlas-tchannel-2,cms-tchannel-1,cms-tchannel-2}.
Furthermore,  ATLAS has found evidence for $\Wt$ associated
production~\cite{atlas-tW}, followed recently by an observation at the
CMS experiment~\cite{cms-tW}. All measurements are in agreement with
SM predictions~\cite{schannel-kidonakis,tchannel-kidonakis}.

%%%%%%%%%%%%%%%%%%%%%%%%%%%
%% 
%% Measurement overview and selections
%% 
%%%%%%%%%%%%%%%%%%%%%%%%%%%

In this Letter, we report final combinations of single-top-quark cross 
section measurements from analyses performed by the
CDF~\cite{cdf_channels} and D0~\cite{d0_schannel} Collaborations using
up to $9.7$\;fb$\rm ^{-1}$ of integrated luminosity per experiment. In
particular, we present a combined $t$-channel cross section, a
combined two-dimensional measurement of the $s$- vs $t$-channel
cross sections, and a combination of the ($s+t$)-channel cross
sections. The combination is obtained by collecting the inputs from
both experiments and reperforming the statistical analysis. This
approach allows for a tighter constraint on the systematic
uncertainties that are common to both experiments, leading to a higher
precision than that achievable from averaging the individual
results. Here, we do not include the combination of the $s$-channel
cross-section measurements, which was reported in
Ref.~\cite{tev_schannel}. We also measure the magnitude of the CKM
matrix element $V_{tb}$ with no assumptions on the number of quark
flavors.

The CDF and D0 detectors are large solenoidal magnetic spectrometers
surrounded by projective-tower-geometry calorimeters and muon
detectors~\cite{CDFII,D0II}. The data were selected using a logical OR
of many online selection requirements that preserve high signal
efficiency for offline analysis. Both collaborations analyze events
with a lepton ($\ell =e$ or $\mu$) plus jets and an imbalance in the
total event transverse energy \MET, reconstructed as the negative
vector sum of all significant transverse energies in the calorimeter
cells and the muon transverse momenta subtracting the calorimeter
energy deposition due to muons (\ljets). This topology is consistent 
with single-top-quark decays in which the decay $W$ boson subsequently
decays to $\ell \nu$~\cite{cdf_channels_7.5,d0_schannel}. Events were
selected that contain only one isolated lepton $\ell$ with large
transverse momentum $p_T$, large \MET, and two or three
clusters of energy in the calorimeters (jets) with large $p_T$. One or
two of these jets were required to be identified as emerging from the 
hadronization of a $b$ quark ($b$-tagged jets). Multivariate
techniques were used to discriminate $b$-quark jets from
light-quark and gluon jets~\cite{CDFbtag,D0btag}. Additional selection criteria were
applied to exclude kinematic regions that were difficult to model and
to minimize the background of multiple jets from QCD production (QCD
multijet) in which one jet was misreconstructed as a lepton and
spurious \MET\ arose from mismeasurements~\cite{cdf_channels_7.5,d0_schannel}. 

The other final-state topology, analyzed by the CDF Collaboration,
involves~\MET, jets, and no reconstructed isolated charged leptons
(\MET+jets)~\cite{cdf_channels}. In the CDF \MET+jets analysis,
overlap with the \ljets sample was avoided by vetoing events with
identified leptons.  Large \MET\ was required, and events with either
two or three reconstructed jets were accepted. This additional sample
increased the acceptance for signal events by including those in which
the $W$-boson decay produced a lepton that is either not reconstructed
or not isolated, or a $\tau$ lepton that decayed into hadrons and a
neutrino, which were reconstructed as a third jet.  After the basic
event selection, QCD multijet events dominate the \MET+jets event
sample. To reduce this background, a selection based on an artificial
neural network was optimized to preferentially select signal-like
events~\cite{cdf_channels}.

Events passing the \ljets and \MET+jets selections were separated into
independent channels based on the number of reconstructed jets as well
as on the number and quality of $b$-tagged jets.  Each of the channels
has a different background composition and signal-to-background ratio,
and analyzing them separately enhances the sensitivity to
single-top-quark production by approximately
10\%~\cite{cdf_channels,d0_schannel}.  

Several differences in the properties of $s$- and $t$-channel events
were used to distinguish them from one another.  Events originating
from $t$-channel production typically contain one light-flavor jet at
large pseudorapidity magnitude $|\eta|$, which is useful for
separating them from events associated with $s$-channel production and
other SM background processes. Events from the $s$-channel process are
more likely to yield two $b$ jets within the central region of the detector.

%%%%%%%%%%%%%%%%%%%%%%%%%%%
%% 
%% Modeling of signal and backgrounds
%% 
%%%%%%%%%%%%%%%%%%%%%%%%%%%

Both collaborations used Monte Carlo~(MC) event generators to simulate
kinematic properties of signal and background events, except for
multijet production, which was modeled with data using matrix
methods~\cite{cdf_matrix_method,d0-prd-2008}. 
Using the \textsc{powheg}~\cite{POWHEG2009} generator, CDF modeled
single-top-quark signal events at NLO accuracy in the strong coupling
strength $\alpha _s$. This is different from D0 where the
{\singletop}~\cite{singletop-mcgen} event generator was used, based on
NLO QCD {\comphep} calculations that match the 
kinematic features predicted by other NLO
calculations~\cite{singletop-xsec-sullivan,Campbell:2009ss}. Spin
information in the decays of the top quark and the $W$~boson is
preserved in both \textsc{powheg} and {\singletop}. 

Kinematic properties of background events from processes in which a
$W$ or $Z$ boson is produced in association with jets ($W$+jets or
$Z$+jets) were simulated using the {\alpgen} MC
generator~\cite{alpgen} for the calculation of tree-level matrix elements 
interfaced to {\pythia}~\cite{pythia} for parton showering and
hadronization and using the MLM matrix-element parton-shower matching
scheme~\cite{MLM}. 
Diboson contributions ($WW$, $WZ$,
and $ZZ$) were modeled using {\pythia}~\cite{pythia}. The {\ttbar}
process was modeled using {\pythia} at CDF and {\alpgen} at D0. The
mass of the top quark in simulated events was set to $m_t=172.5$~GeV. 
Higgs-boson processes were modeled using simulated events generated
with {\pythia} for a Higgs boson mass of
$m_H=125$~GeV~\cite{atlas-higgs-mass,cms-higgs-mass1,cms-higgs-mass2}.  
In all of the above cases, {\pythia} was used to model proton remnants
and to simulate the hadronization of all generated partons.  The
presence of additional \ppbar\ interactions was modeled by overlaying
events selected from random beam crossings matching the instantaneous
luminosity profile in the data. All MC events were processed through
{\geant}-based detector simulations~\cite{geant}, and were reconstructed
using the same computer programs as used for data. 

Data were used to normalize $W$-boson production associated with
both light- and heavy-flavor jet contributions in samples enriched in
$W$+jets processes, which have negligible signal
content~\cite{d0_schannel,cdf-prd-2010,cdf_schannel_MET}.  All other
simulated background samples were normalized to their theoretical
cross sections, i.e., {\ttbar} at next-to-next-to-leading order
QCD~\cite{ttbar-xsec}, $Z$+jets and diboson production at NLO
QCD~\cite{mcfm}, and Higgs-boson production including all relevant
higher-order QCD and electroweak corrections~\cite{higgs-xsec}. For
the measurement of $\sigma_t$, the $s$-channel single-top-quark
production sample was considered as background and normalized to the
NLO QCD cross section combined with NNLL
resummations~\cite{schannel-kidonakis}.

%%%%%%%%%%%%%%%%%%%%%%%%%%%
%% 
%% Combination method
%% 
%%%%%%%%%%%%%%%%%%%%%%%%%%%

Multivariate discriminants were optimized to separate signal events from
large background contributions. To combine the results from the two
experiments, we use the $s$- and $t$-channel discriminants from the
CDF~\cite{cdf_schannel} and D0~\cite{d0_schannel} single-top-quark
measurements. We perform a likelihood fit to the binned distribution
of the final discriminants. We combine the various channels of the
different analyses from each experiment by taking the product of their
likelihoods and simultaneously varying the correlated uncertainties
and by comparing data to the predictions for each contributing signal
and background process. Using a Bayesian statistical
analysis~\cite{bayes-limits}, we then derive combined Tevatron cross
section measurements, taking the prior density for the signal cross
sections to be uniform for non-negative cross sections.

%%%%%%%%%%%%%%%%%%%%%%%%%%%
%% 
%% Systematic uncertainties
%% 
%%%%%%%%%%%%%%%%%%%%%%%%%%%

For the sources of uncertainties we follow Ref.~\cite{tev_schannel}.
We consider the following systematic uncertainties: the integrated
luminosity from detector-specific sources and from the inelastic and
diffractive cross sections.  We also consider systematic uncertainties
on the signal modeling, the simulation of background, data-based
methods to estimate background, detector modeling, $b$-jet tagging, and the
measurement of the jet-energy scale. Table I of
Ref.~\cite{tev_schannel} summarizes the categories that contribute to
the uncertainties on the shape of the output of the multivariate
discriminants distributions and the range of uncertainties applied to
the predicted normalizations for signal and background contributions.
Reference~\cite{tev_schannel} gives the sources of systematic
uncertainty common to measurements of both collaborations that are
assumed to be fully correlated, and lists uncertainties that are
assumed to be uncorrelated. The dependence of the results on these
correlation assumptions is negligible. 

A two-dimensional (2D) posterior-probability density is constructed as
a function of $\sigma_s$ and $\sigma_t$ in analogy to the
one-dimensional (1D) posterior probability described in Ref.~\cite{tev_schannel}. 
The measured cross section is quoted as the value at the position of
the maximum, and the 68\% probability contour defines the measurement
uncertainty.

%%%%%%%%%%%%%%%%%%%%%%%%%%%
%% 
%% Results
%% 
%%%%%%%%%%%%%%%%%%%%%%%%%%%

Figure~\ref{fig:TevtbtqbSubtrLog} shows the distribution of the
mean values from the discriminants sorted by the $s$-channel minus
$t$-channel expected signal contributions divided by the background 
expectation, $(s-t)/b$. An entry in the histogram corresponds to a
collection of bins with similar ratio $(s-t)/b$. The value on the
horizontal axis is given by the mean discriminant for those bins. The
vertical axis gives the number of events in those bins.
We show the data, the SM predictions for the $s$-  and
$t$-channel processes, and the predicted backgrounds separated by
source. 
The distribution for large negative values is dominated by the content
of the bins that show a higher $t$-channel contribution, while large
positive values are dominated by the content of the bins with a higher
$s$-channel contribution. The abscissa extends to larger negative
values since we expect more $t$-channel events than $s$-channel events
and the separation from background is better for $t$-channel events
than for $s$-channel events. The region corresponding to discriminant
values near zero is dominated by the background. 

%%%%%%%%%%%%%%%%%%%%%%%%%%%%%%%%%%%%%%%%%%%%%%%%%%%%
\begin{figure}[!h!tbp]
\begin{center}
\includegraphics[width=0.48\textwidth]{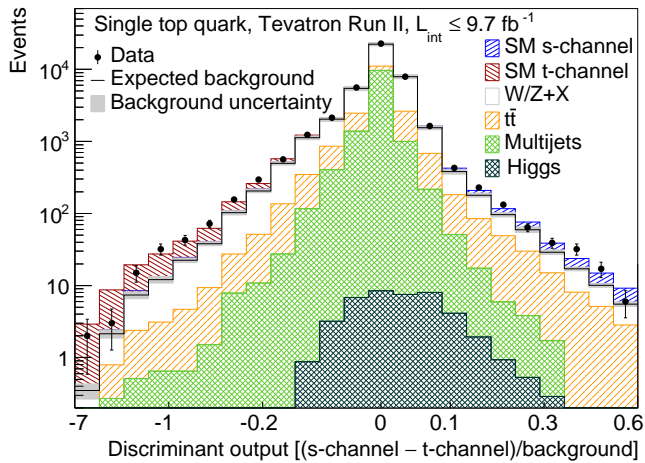}
\caption{Distribution of the mean discriminants for bins
  with similar ratios of ($s$-channel$-$$t$-channel) signals divided by
  background yields. The data,
  predicted SM $s$- and $t$-channel yields, and expected background
  are displayed. The total expected background (black solid line) is
  shown with its uncertainty (gray shaded band). A nonlinear scale is
  used on the abscissa to better display the range of the discriminant
  output values.}
\label{fig:TevtbtqbSubtrLog}
\end{center}
\end{figure}
%%%%%%%%%%%%%%%%%%%%%%%%%%%%%%%%%%%%%%%%%%%%%%%%%%%%

Figure~\ref{fig:tevposterior} presents the resulting 2D posterior
probability distribution as a function of $\sigma_t$ and $\sigma_s$.
%
%%%%%%%%%%%%%%%%%%%%%%%%%%%%%%%%%%%%%%%%%%%%%%%%%%%%
\begin{figure}[!h!tbp]
\begin{center}
\includegraphics[width=0.48\textwidth]{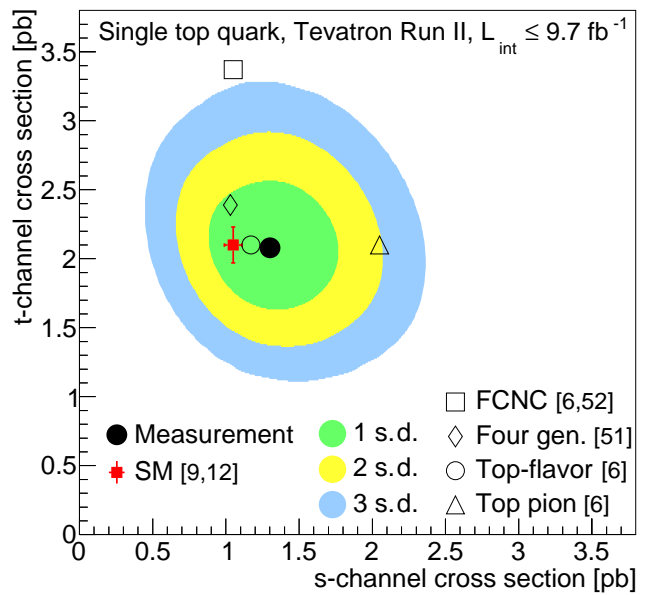}
\caption{Two-dimensional posterior probability as a function of
  $\sigma_t$ and $\sigma_s$ with one s.\ d. (68\% C.L.), two
  s.d. (95\% C.L.), and three s.\ d. (99.7\% C.L.) probability
  contours for the combination of the CDF and D0 analysis channels
  compared with the NLO+NNLL theoretical prediction of the
  SM~\cite{schannel-kidonakis,tchannel-kidonakis}. Several BSM
  predictions are shown, a model with four quark families with
  top-to-strange quark coupling $|V_{ts}| = 0.2$~\cite{FourthGen2}, a
  top-flavor model with new heavy bosons with mass $m_x = 
  1$\;TeV~\cite{Tait:2000sh}, a model of charged top pions with mass
  $m_{\pi^{\pm}} = 250$\;GeV~\cite{Tait:2000sh}, and a model with
  flavor-changing neutral currents with a 0.036 coupling
  $\kappa_u/\Lambda$ between up quark, top quark, and
  gluon~\cite{Tait:2000sh,d0-fcnc}.}
\label{fig:tevposterior}
\end{center}
\end{figure}
%%%%%%%%%%%%%%%%%%%%%%%%%%%%%%%%%%%%%%%%%%%%%%%%%%%%
%
The value and uncertainty in the individual cross sections are derived
through the 1D posterior probability functions obtained by integrating
the 2D posterior probability over the other variable. The most
probable value of $\sigma_t$ is $2.25 ^{+0.29}_{-0.31}$~pb. The
measurement of $\sigma_{s+t}$ is performed without making assumptions
on the ratio of $\sigma_s/\sigma_t$ by forming a 2D posterior
probability density distribution of $\sigma_{s+t}$ versus $\sigma_t$
and then integrating over all possible values of $\sigma_t$ to extract
the 1D estimate of $\sigma_{s+t}$. The combined cross section is
$\sigma_{s+t} = 3.30^{+0.52}_{-0.40}$~pb. The total expected
uncertainty on $\sigma_{s+t}$ is 13\%, the expected uncertainty
without considering systematic uncertainties is 8\%, and the
expected systematic uncertainty is 10\%. The systematic
uncertainty from the limited precision of top-quark mass measurements
is negligible~\cite{cdf-prd-2010,d0_schannel}. Figure~\ref{fig:tevposterior} 
also shows the expectation from several beyond the SM (BSM) models. 
Figure~\ref{fig:tevxsec} shows the
individual~\cite{cdf_channels,d0_schannel} and combined (this Letter)
measurements of the $t$- and ($s+t$)-channel cross sections including
previous measurements of the
individual~\cite{cdf_schannel,d0_schannel} and
combined~\cite{tev_schannel} $s$-channel cross sections. All
measurements are consistent with SM predictions.
%
%%%%%%%%%%%%%%%%%%%%%%%%%%%%%%%%%%%%%%%%%%%%%%%%%%%%
\begin{figure}[!h!tbp]
\begin{center}
\includegraphics[width=0.48\textwidth]{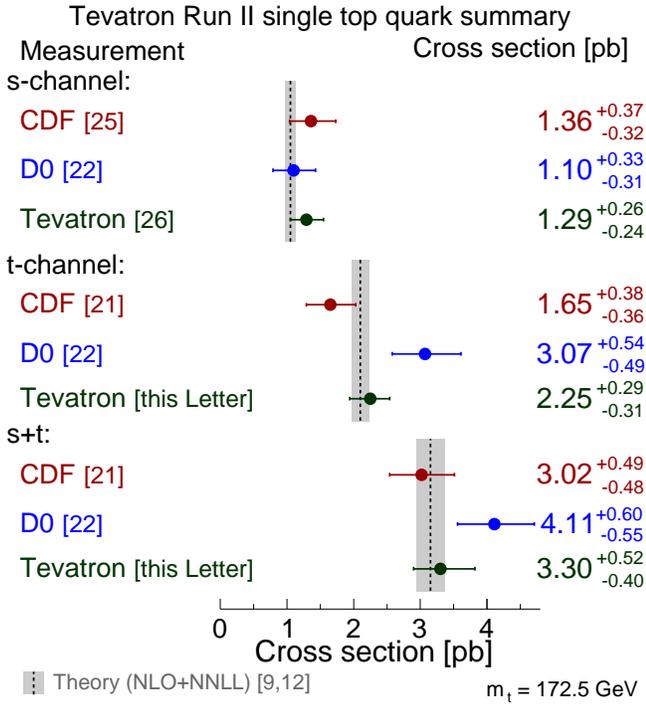}
\caption{Measured single-top-quark production cross
  sections from the CDF and D0 Collaborations in different production
  channels and the Tevatron combinations of these analyses compared
  with the NLO+NNLL theoretical
  prediction~\cite{schannel-kidonakis,tchannel-kidonakis}.}
\label{fig:tevxsec}
\end{center}
\end{figure}
%%%%%%%%%%%%%%%%%%%%%%%%%%%%%%%%%%%%%%%%%%%%%%%%%%%%

The SM single-top-quark production cross section is directly sensitive
to the square of the CKM matrix element
$V_{tb}$~\cite{tchannel-kidonakis,schannel-kidonakis}, thus  
providing a measurement of $|V_{tb}|$ without any assumption
on the number of quark families or the unitarity of the CKM
matrix~\cite{d0-prd-2008}.  
We extract $|V_{tb}|$ assuming that top quarks decay exclusively to
$Wb$ final states.  

We start with the multivariate discriminants for the $s$ and $t$
channels for each experiment and form a Bayesian posterior probability
density for $|V_{tb}|^2$ assuming a uniform-prior probability
distribution in the region $[0,\infty]$ corresponding to a uniform
prior density of the signal cross section. Additionally, the
uncertainties on the SM predictions for the $s$- and $t$-channel cross
sections~\cite{schannel-kidonakis,tchannel-kidonakis} are
considered. The resulting posterior probability distribution for
$|V_{tb}|^2$ is presented in Fig.~\ref{fig:TevVtb}. 
%
%%%%%%%%%%%%%%%%%%%%%%%%%%%%%%%%%%%%%%%%%%%%%%%%%%%%
\begin{figure}[!h!tbp]
\begin{center}
\includegraphics[width=0.48\textwidth]{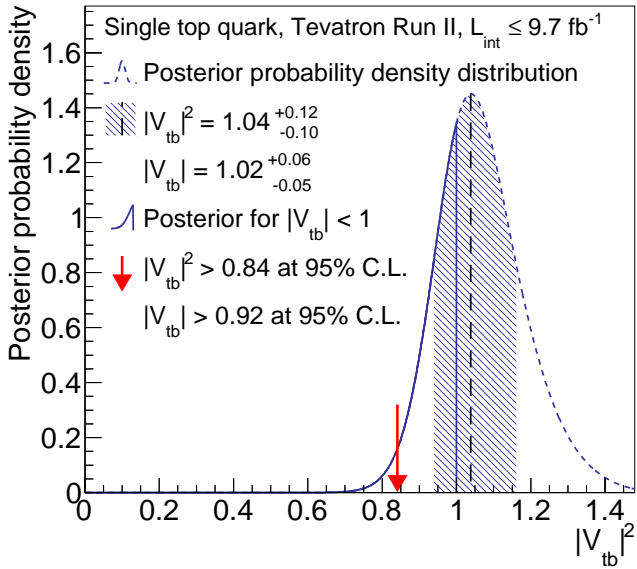}
\caption{Posterior probability distribution as a
  function of $|V_{tb}|^2$ for the combination of CDF and D0 analysis
  channels. The arrow indicates the allowed values of $|V_{tb}|^2$
  corresponding to the limit of $|V_{tb}| > 0.92$ at the $95\%$~C.L.} 
\label{fig:TevVtb}
\end{center}
\end{figure}
%%%%%%%%%%%%%%%%%%%%%%%%%%%%%%%%%%%%%%%%%%%%%%%%%%%%
%
We obtain $|V_{tb}| = 1.02^{+0.06}_{-0.05}$. If we restrict the prior
to the SM region [0,1], we extract a limit of $|V_{tb}| > 0.92$ at the
$95\%$~C.L.

%%%%%%%%%%%%%%%%%%%%%%%%%%%
%% 
%% Summary
%% 
%%%%%%%%%%%%%%%%%%%%%%%%%%%

In summary, using $p\bar{p}$ collision samples corresponding to an
integrated luminosity of up to 9.7\;fb$^{-1}$ per experiment, we
report the final combination of single-top-quark production cross
sections from CDF and D0 measurements assuming $m_t=172.5$\;GeV. The
cross section for $t$-channel production is found to be  
\[\sigma_t = 2.25 ^{+0.29}_{-0.31}\;\rm pb .\] Without assuming the SM
value for the relative $s$- and $t$-channel contributions, the total
single-top-quark production cross section is \[\sigma_{s+t} =
3.30^{+0.52}_{-0.40}\;\rm pb.\] Together with the combined $s$-channel
cross section~\cite{tev_schannel}, this completes single-top-quark
cross-section measurements accessible at the Tevatron. All
measurements are consistent with SM
predictions~\cite{schannel-kidonakis,tchannel-kidonakis}. Finally, we
extract a direct limit on the CKM matrix element of $|V_{tb}| > 0.92$
at the $95\%$~C.L. As a result, there is no indication of sources of
new physics beyond the SM in the measured strength of the $Wtb$
coupling.

%---------------------------------------------------------------------

%%%%%%%%%%%%%%%%%%%%%%%%%%%
\section*{Acknowledgments}
We thank the Fermilab staff and technical staffs of the participating
institutions for their vital contributions. We acknowledge support
from the Department of Energy and the National Science Foundation
(U.S.A.), the Australian Research Council (Australia), the National
Council for the Development of Science and Technology and the Carlos
Chagas Filho Foundation for the Support of Research in the State of
Rio de Janeiro (Brazil), the Natural Sciences and Engineering Research
Council (Canada), the China Academy of Sciences, the National Natural
Science Foundation of China, and the National Science Council of the
Republic of China (China), the Administrative Department of Science,
Technology and Innovation (Colombia), the Ministry of Education, Youth
and Sports (Czech Republic), the Academy of Finland, the Alternative
Energies and Atomic Energy Commission and the National Center for
Scientific Research/National Institute of Nuclear and Particle Physics
(France), the Bundesministerium f{\"u}r Bildung und Forschung (Federal
Ministry of Education and Research) and the Deutsche
Forschungsgemeinschaft (German Research Foundation) (Germany), the
Department of Atomic Energy and Department of Science and Technology
(India), the Science Foundation Ireland (Ireland), the National
Institute for Nuclear Physics (Italy), the Ministry of Education,
Culture, Sports, Science and Technology (Japan), the Korean World
Class University Program and the National Research Foundation of Korea
(Korea), the National Council of Science and Technology (Mexico), the
Foundation for Fundamental Research on Matter (Netherlands), the
Ministry of Education and Science of the Russian Federation, the
National Research Center ``Kurchatov Institute'' of the Russian
Federation, and the Russian Foundation for Basic Research (Russia),
the Slovak R\&D Agency (Slovakia), the Ministry of Science and
Innovation, and the Consolider-Ingenio 2010 Program (Spain), the
Swedish Research Council (Sweden), the Swiss National Science
Foundation (Switzerland), the Ministry of Education and Science of
Ukraine (Ukraine), the Science and Technology Facilities Council and
the The Royal Society (United Kingdom), the A.P. Sloan Foundation
(U.S.A.), and the European Union community Marie Curie Fellowship
Contract No. 302103.
%

%%%%%%%%%%%%%%%%%%%%%%%%%%%

%%%%%%%%%%%%%%%%%%%%%%%%%%%

\end{document}

%---------------------------------------------------------------------
%---------------------------------------------------------------------